\documentclass{JHEP3}

\usepackage{amsfonts,amsbsy,eucal,mathrsfs,bbm,xy}
\xyoption{all}

\usepackage{graphicx}
\usepackage{amssymb}
\usepackage{epstopdf}
\DeclareSymbolFont{lett}{U}{eur}{m}{n}
\SetSymbolFont{lett}{bold}{U}{eur}{b}{n}
\DeclareMathSymbol\LL \mathord{lett}{"03}
\DeclareMathSymbol\dd \mathord{lett}{"0E}

\catcode`\@=11 

\global\newcount\nsecno \global\nsecno=0
\global\newcount\meqno \global\meqno=1
\def\newsec#1{\global\advance\nsecno by1
\eqnres@t
\section{#1}}
\def\eqnres@t{\xdef\nsecsym{\the\nsecno.}\global\meqno=1}
\def\sequentialequations{\def\eqnres@t{\bigbreak}}\xdef\nsecsym{}

\def\draftmode{\message{ DRAFTMODE }
\writelabels

{\count255=\time\divide\count255 by 60 \xdef\hourmin{\number\count255}
\multiply\count255 by-60\advance\count255 by\time
\xdef\hourmin{\hourmin:\ifnum\count255<10 0\fi\the\count255}}}
\def\nolabels{\def\wrlabeL##1{}\def\eqlabeL##1{}\def\reflabeL##1{}}
\def\writelabels{\def\wrlabeL##1{\leavevmode\vadjust{\rlap{\smash%
{\line{{\escapechar=` \hfill\rlap{\tt\hskip.03in\string##1}}}}}}}%
\def\eqlabeL##1{{\escapechar-1\rlap{\tt\hskip.05in\string##1}}}%
\def\reflabeL##1{\noexpand\llap{\noexpand\sevenrm\string\string\string##1}
}}

\nolabels

\def\eqn#1#2{
\xdef #1{(\nsecsym\the\meqno)}
\global\advance\meqno by1
$$#2\eqno#1\eqlabeL#1
$$}

\def\eqalign#1{\null\,\vcenter{\openup\jot\m@th
\ialign{\strut\hfil$\displaystyle{##}$&$\displaystyle{{}##}$\hfil
\crcr#1\crcr}}\,}

\def\foot#1{\footnote{#1}}

\catcode`\@=12 
\def\a{\alpha}
\def\b{\beta}
\def\c{\chi}
\def\d{\delta}  
\def\e{\varepsilon} 
  
\def\g{\gamma}  \def\G{\Gamma}

\def\l{\lambda}

\def\r{\rho}

  \def\O{\Omega}
\def\p{\psi}  
\def\s{\sigma}  \def\S{\Sigma}
  
\def\t{\tau}
\def\w{\varphi}

\def\CM{{\cal M}}
\def\CN{{\cal N}}
\def\CO{{\cal O}}

\def\CT{{\CMcal T}}

\def\C{\mathbb{C}}

\def\mh{\mathfrak{h}}

\def\mt{\mathfrak{t}}

\def\mD{\mathfrak{D}}

\def\sL{\mathscr{L}}

\def\bos#1{\boldsymbol{#1}}

\def\roughly#1{\raise.3ex\hbox{$#1$\kern-.75em\lower1ex\hbox{$\sim$}}}

\def\rd{\partial}

\def\fr#1#2{{\textstyle{#1\over#2}}}
\def\Fr#1#2{{#1\over#2}}
\def\ato#1{{\buildrel #1\over\longrightarrow}}

\def\pr{\prime}

\def\maprd{\rlap{\lower.3ex\hbox{$\scriptstyle\bs_+$}}\searrow}
\def\mapld{\swarrow\!\!\!\rlap{\lower.3ex\hbox{$\scriptstyle\bs_-$}}}

\vspace{.2in}

\title{Topological Sigma B Model in 4-Dimensions}

\author{Hyun-Keun Jun \\
Department of Mathematics, Kyungpook University,
Daegu 702-701, Korea.
 \\ E-mail:\email{hyunkeun@knu.ac.kr}}
\author{Jae-Suk Park\thanks{This work was supported by 
KOSEF Interdisciplinary Research Grant No. R01-2006-000-10638-0.} \\
Department of Mathematics, Yonsei University, Seoul 120-749, Korea.\\ 
E-Mail:\email{jaesuk@yonsei.ac.kr}}


\abstract{
We propose a 4-dimensional version of topological sigma B-model, governing maps from a smooth 
compact $4$-manifold
$M$ to a Calabi-Yau target manifold $X$.  The theory depends on complex structure of $X$, while
is independent  of K\"ahler metric of $X$. The theory is also a 4-dimensional topological field theory in the sense 
that the theory is independent of variation of  Riemannian metric of the source $4$-manifold $M$, potentially leading
to new smooth invariant of $4$-manifolds.
We argue that the theory also comes with a topological family parametrized by the extended moduli space
of complex structures.}

\preprint{}

\begin{document}

\newsec{Introduction}

The purpose of this paper is to propose a $4$-dimensional topological sigma model,
which governs maps from a smooth oriented compact $4$-manifold  $M$ to a Calabi-Yau manifold
$X$.  Our model is a topological theory in $4$-dimensions in the sense that the variation of action functional
with respect to  Riemannian metric of the source $4$-manifold $M$ is BRST exact. 
Analogous to the topological string B-model in $2$-dimensions \cite{W1} our model does not
depend on variation of K\"ahler structures of the target Calabi-Yau space $X$, while depending on  variation
of complex structures of $X$. We shall also argue that there should be a family of topological theory parametrized
by the extended moduli space of complex structures of $X$ similarly to the topological string B model \cite{W1,BK,P1}. 

Our theory may lead to potentially new smooth invariant, a family of invariants
parametrized by the moduli space of complex structures of Calabi-Yau manifold, of $4$-manifold. 
Our theory  has  many properties similar to those of Donaldson-Witten theory \cite{W2}, while a perfect analogy
between the two theories seems require the target space to admit a holomorphic symplectic structure.
This issue will be pursued in a forthcoming paper \cite{JP1}.

This paper is organized as follows;  In section $2$ we  briefly review some basic properties of 
the Atiyah class of holomorphic tangent space of Calabi-Yau space 
and construct the Lagrangian of our model. Then we show that the variations of our Lagrangian with respect
to infinitesimal variations of both Riemannian metric of the source $4$-manifold $M$ and K\"ahler metric of 
the target Calabi-Yau manifold $X$ are BRST exact. We also 
show that path integrals of the model are localized to the space of constant maps and construct topological observables.  
In section $3$ we study infinitesimal deformations of the theory and argue that  
there should be a family of theory parametrized by the extended moduli space of complex structures of $X$.

\newsec{Construction of Our Model}

Our model will be a four dimensional sigma model, which is a theory of maps from
a smooth oriented Riemannian $4$-manifold $M$ to a compact Calabi-Yau manifold
$X$.   To define our Lagrangian, we need to pick a Riemannian metric $h$ on $M$, a complex structure on $X$
as well as a K\"ahler metric $g$ on $X$, while the theory will depend only 
on the complex structure on $X$ - that is, independent
to infinitesimal changes in both $h$ and $g$.  This feature is similar to that of topological string B model.
We begin with setting up notations and conventions.

Let $M$ be a smooth oriented compact Riemannian $4$-manifold. 
We denote the space of differential $r$-forms on $M$ by $\O^r(M)$ 
and  the exterior derivative on $M$ by $d$, i.e., $d:\O^r(M)\rightarrow \O^{r+1}(M)$
satisfying $d^2 =0$. We pick a Riemannian metric $h$ on $M$ and denote $*$ the corresponding Hodge star operator,
so that $*: \O^r(M) \longrightarrow \O^{4-r}(M)$ and $*^2=1$. Then the space $\O^2(M)$ of $2$-forms
is decomposed into
the self-dual and anti-self-dual parts;
\eqn\aaa{
\O^2(M) = \O^2_+(M)\oplus \O^2_-(M).
}
For a two form $a$ we denote $a^+$ the projection to the self-dual part of $a$, i.e., $* a^+= a^+$.

Let $X$ be a complex $n$-dimensional compact Calabi-Yau manifold - a K\"{a}her manifold with $c_1(X)=0$. 
Let $T_\C X$ denote the complexified tangent bundle to $X$.
We pick a complex structure $J$, $J^2=-1$, on $X$ and extend $J$ to a complex endomorphism of $T_\C X$
leading to the eigen-space decompositions $T_\C X = \CT \oplus \overline\CT$, where $\CT$
is the holomorphic tangent  bundle of rank $n$. Similarly  the complexified cotangent bundle $T^*_\C X$
has the decomposition $T^*_\C X = \CT^*\oplus \overline\CT^*$, where $\CT^*$
is the holomorphic cotangent bundle. 
We denote the space of differential $(p,q)$-forms
on $X$ by $\O^{p,q}(X) = \O^{0,q}(X,\wedge^p \CT^*)$. The exterior derivative $d$ is decomposed
as $d=\rd + \bar\rd$, where $\bar\rd : \O^{p,q}(X) \longrightarrow \O^{p+1,q}(X)$ and $\bar\rd^2=0$. 
We also consider the space $\O^{0,q}(X,\wedge^p \CT)$
of $(0,q)$-form on $X$ valued in the section of $\wedge^p \CT$.
Let $\O$ be a choice of no-where vanishing holomorphic $n$-form on $X$.
Then we have the familiar isomorphism
\eqn\aab{
\O^{0,q}(X,\wedge^p \CT) \simeq \O^{n-p,q}(X) = \O^{0,q}(X,\wedge^{n-p}\CT^*).
}

Pick a K\"ahler metric $g$ and let $\widetilde \nabla$ be the unique $g$-connection of $\CT$ such that
the $(0,1)$-part of connection is $\bar\rd$.
Let $\nabla$ be the canonical $(1,0)$-part of the metric connection on $\CT$
such that the $(2,0)$-part of the curvature $2$-form vanishes;
\eqn\aac{
[\nabla, \nabla] =0.
}
Then the curvature $R$ of  the total connection $\widetilde\nabla = \nabla +\bar\rd$ 
is an $(1,1)$-form on $X$
valued in $End(\CT)$;
\eqn\aad{
R =[\bar\rd,\nabla] \in\O^{1,1}(End(\CT)).
}  
Pick a local holomorphic coordinates $\{u^i\}$, $i=1,2,\ldots n$, and their complex conjugates $\{u^{\bar i}\}$. 
Denoting the  K\"ahler metric $g$  in terms of local coordinates by $g_{i\bar j}$, 
the associated connection $(1,0)$-form is 
\eqn\aaf{
\G^i{}_{jk}dz^k:=  g^{i\bar\ell}\rd_j g_{k\bar \ell} du^k,
}
where $\rd_j= \Fr{\rd}{\rd u^j}$.
Note that $\G^i{}_{jk}=\G^i{}_{kj}$.
Then the curvature $R$ of $\CT$ in components is given by
\eqn\aag{
R^i{}_{jk\bar\ell} = \rd_{\bar \ell}\G^{i}{}_{jk}.
}
It follows that $R^i{}_{jk\bar\ell} =R^i{}_{kj\bar\ell}$, so that $R \in \O^{0,1}(X, Hom(S^2\CT, \CT))$.
It is also obvious that $\bar\rd R =0$. On the other hand,  $\nabla R \neq 0$ in general,
leading to  the
"higher curvature" tensor $S \in \O^{0,1}(X, Hom(S^2\CT\otimes \CT, \CT)$
defined by $S:=[\nabla, R]$;
\eqn\aah{
S^{i}{}_{mjk\bar \ell}:= \nabla_m R^{i}{}_{jk\bar \ell}.
}
{}From the relation \aac, it follows that $S^{i}{}_{mjk\bar \ell}$ is actually symmetric in all three indices $m,j,k$.
We recall that $\bar\rd R =0$ due to the definition \aad, which condition in components is
\eqn\aai{
\rd_{\bar m}R^i{}_{jk\bar \ell}\p^{\bar m}\p^{\bar \ell} =0,
} 
where we  used anti-commuting variables  $\p^{\bar i}$,
for a convenience, and the Einstein summation conventions (to be used throughout this paper). 
On the other hand $S$ is not $\bar\rd$-closed in general. We have
\eqn\aaj{
\eqalign{
\rd_{\bar n}\left(S^i{}_{mjk\bar\ell}\right)t^j t^k \p^{\bar n}\p^{\bar\ell}
&= \left(R^i{}_{pm\bar n}R^p{}_{jk\bar \ell} + 2 R^i{}_{pj\bar n}R^{p}_{\bar k m\bar\ell}\right)
t^j t^k \p^{\bar n}\p^{\bar \ell},
}
} 
where we further used  commuting variables $t^i$, for a convenience. 
We remark that the identity \aaj\ was derived in the appendix of the paper \cite{RW}.
We also remark that the curvature $R$ and the higher curvature $S$
are the Dolbeault representatives of the Atiyah class and a higher Atiyah class of $\CT$, which are
the lowest parts of the sequence of higher Atiyah classes \cite{K, Ka}. 

\subsection{Global Fermionic Symmetry and Lagrangian}

Our model is a theory of maps $\phi: M \rightarrow X$ from the Riemanian $4$-manifold
$M$ to the Calabi-Yau manifold $X$, as are specified in the above,
with the choices of a Riemannian metric on $M$, a complex structure on $X$ and a K\"aher metric  on $X$. 
A map $\phi: M \rightarrow X$ can be described locally by functions $u^i(x)$ and their complex
conjugates $u^{\bar i}(x)$, where $x$ denotes a a point in $M$ and the $u^i$ corresponds to holomorphic
coordinates on $X$. We shall denote the corresponding bosonic fields by $u^i$ and $u^{\bar i}$, which
are $0$-forms on $M$. 
We introduce an integral valued quantum number called ghost number
$U$ and assign  $U=0$ to the fields $(u^i, u^{\bar i})$.
We introduce an anti-commuting $0$-form field $\p^{\bar i}$  of $U=1$, which is a section of $\phi^*(\overline\CT)$,
the pullback of anti-holomorphic tangent bundle of $X$ to $M$. We also introduce another
anti-commuting $0$-form field $\eta_i$ with $U=3$, a section of $\phi^*(\CT^*)$. 
We further introduce an anti-commuting $1$-form field $\r^i= \r^i_\a dx^\a$ of $U=-1$ and
a commuting {\it self-dual} $2$-form field $H^i = \Fr{1}{2}H^i_{\a\b}dx^\a\wedge dx^\b$ of $U=-2$
- the both fields are sections of $\phi^*(\CT)$. 
In the above and throughout this paper $\a,\b=1,2,3,4$ are tangent indices to $M$ and $\{x^\a\}$ denotes coordinates on $M$.
We also introduce a commuting  $1$-form field $A_i=A_{i\a}dx^\a$ of $U=2$ and an anti-commuting 
{\it self-dual} $2$-form field $\chi_i=\Fr{1}{2}\chi_{i\a\b}dx^\a\wedge dx^\b$ of $U=1$ -the
both fields are sections of $\phi^*(\CT^*)$.
Finally we require two more $0$-form fields $(\bar \eta^i, v^i)$  of $U=(-3,-2)$, which are  sections
of $\phi^*(\CT^i)$. A field with even ghost number $U$ is
commuting and a field with odd ghost number $U$ is anti-commuting.
We summarize  all the fields in the following table;
\begin{table}[h]\label{table1}
\begin{center}
\begin{tabular}{|c|c|c|c|c|c|c|c|}
   \hline
   form degree$\setminus$ghost number &-3 & -2 & -1& 0 & 1& 2 & 3\\
   \hline
   $\Omega^0(M)$  & $\bar{\eta}^i$ & $v^i$ & &$ u^i, u^{\bar i}$ &$\psi^{\bar i}$& &$\eta_i$ \\
   \hline
   $\Omega^1(M)$   &&& $\rho^i$&&& $w_i$ &  \\
   \hline
   $\Omega^2_{+}(M)$&&$H^i$ &&& $\chi_i$ &&  \\
   \hline
\end{tabular}
\end{center}
\caption{The form degree on $M$ and the ghost numbers of the various fields}
\end{table}
\newpage
For a later purpose we state that the indices tangent to $M$ are raised and lowered by a Riemannian metric $h_{\a\b}$ on $M$.
Then the self-duality conditions of the two $2$-form fields $\chi_i=\fr{1}{2}\chi_{i\a\b}dx^\a\wedge dx^\b$
and $H^i=\fr{1}{2}H^i_{\a\b}dx^\a\wedge dx^\b$ are
\eqn\zabn{
\eqalign{
 H^{i}_{\a\b}&=\frac{1}{2}\e_{\a\b\g\d}h^{\a\g'}h^{\d\d'} H_{\g'\d'}^{i}
 ,\cr
\chi_{i\a\b}&=\frac{1}{2}\e_{\a\b\g\d}h^{\g\g'}h^{\d\d'}\chi_{i\g'\d'}.
}
}

Now we state conventions on covariant derivatives.
A connection on $\phi^*(\overline\CT)$ is obtained by pulling back the metric connection of $\overline \CT$ from $X$ to
$M$, giving the covariant derivative of $\p^{\bar i}$
\eqn\abba{
\overline D \p^{\bar i} = d\p^{\bar i} + du^{\bar j}\G^{\bar i}{}_{\bar j\bar k}\p^{\bar k},
}
where $\G^{\bar i}{}_{\bar j\bar k} = g^{\bar i \ell}\rd_{\bar j}g_{\ell\bar k}$,
and similarly for any other section of $\phi^*(\overline\CT)$.
A connection on $\phi^*(\CT^*)$ is obtained by pulling back the metric connection of $\CT^*$ from $X$ to
$M$, giving the covariant derivative of $\eta_{i}$
\eqn\abbb{
\eqalign{
D \eta_i &= d\eta_i - du^{j}\G^{k}{}_{ij}\eta_k,\cr
}
}
and similarly for any other section of $\phi^*(\CT^*)$.
We also introduce a twisted covariant derivative $\mathfrak{D}$ acting on the various sections of 
$\phi^*(\CT^*)$ by the formula
\eqn\split{
\eqalign{
\mathfrak{D}\eta_{i}&=D\eta_{i}-R_{ij\bar\ell}^{k}\psi^{\bar\ell}\rho^{j}\eta_{k},\cr
\mathfrak{D} w_{i}&=D w_{i}-R_{ij\bar\ell}^{k}\psi^{\bar\ell}\rho^{j}\land
 w_{k}.
}
}
As the rationale behind the above definition, we first recall  that 
the field $\r^i$ is an anti-commuting $1$-form on $M$. The
twisted covariant derivative $\mathfrak{D}$ can be compared with the usual gauge
covariant derivative in Yang-Mills theory on $M$, where the
exterior derivative is replaced with $D$, the gauge $1$-form connection
is replaced with $\r^i$ and the structure constant is replaced with
$R^k{}_{ij\bar\ell}\p^{\bar \ell}$, and the Bianchi identity is replaced with the identity  in \aaj.
We should remark that the closely related thing happens in
the Rozanky-Witten-Kontsevich-Kapranov theory, which is
a $3$-dimensional topological sigma model with a holomorphic
symplectic manifold as the target space.

Our model has a global fermionic symmetry. Before stating its transformation laws, 
it is convenient to divide all the fields into four kinds of multiplets;
\begin{itemize}

\item
$(u^i, \r^i, H^i)$ with $U=(0,-1,-2)$ in $\left(\O^0(M), \O^1(M), \O^2_+(M)\right)$.
\item
$(u^{\bar i}, \p^{\bar i})$ with $U=(0,1)$ both in $\O^0(M)$.

\item
$(\eta_i, w_i, \chi_i)$ with $U=(3,2,1)$ in $\left(\O^0(M), \O^1(M), \O^2_+(M)\right)$.
\item
$(\bar\eta^{ i}, v^{i})$ with $U=(-3,-2)$ both in $\O^0(M)$.
\end{itemize}

We first postulate the fermionic  transformation laws for $(u^i, \r^i, H^i)$ and $(u^{\bar i}, \p^{\bar i})$
\eqn\abd{
\eqalign{
Q u^i &=0
,\cr
Q \r^i &=du^i
,\cr
Q H^i  &=- \left(D \r^i + \fr{1}{2}R^i{}_{jk\bar\ell}\r^j\wedge\r^k \p^{\bar \ell}\right)^+
,\cr
Q u^{\bar i} &=\p^{\bar i},\cr
Q \p^{\bar i} &= 0.\cr
}
}
We also postulate
that
\eqn\abda{
\eqalign{
Q\eta_i =&0
,\cr
Q w_i = &-\mathfrak{D} \eta_i
,\cr
Q \c_i  =&\left(\mathfrak{D}  w_i\right)^+ 
	-R^k{}_{ij\bar \ell}  H^j \p^{\bar \ell}\eta_k
	\cr
	&
	-R^{k}{}_{ij\bar \ell} \left(du^{\bar \ell}\wedge\r^j\right)^+ \eta_k 
		\cr &
	+\fr{1}{2}S^{n}{}_{ijk\bar \ell}\left(\r^j\wedge\r^k\right)^+ \p^{\bar \ell}\eta_n 
,\cr
Q\bar\eta^i =&v^i,\cr
Q v^i =&0,\cr
}
}
for $(\eta_i, w_i, \chi_i)$ and $(\bar\eta^{ i}, v^{i})$.
The global fermionic supercharge $Q$ carries $U=1$ and
satisfies $Q^2=0$. The  property $Q^2=0$  may be checked explicitly - the  only non-trivial cases are
for $Q^2 H^i$ and  $Q^2 \chi_i$ and a little tedious but straightforward computations show that
$$
\eqalign{
Q^2 H^i =& \frac{1}{2}\rd_{\bar n}R^{i}_{jk\bar\ell}\psi^{\bar n}\psi^{\bar\ell}\rho^j\land\rho^k,\cr
Q^2 \chi_i=&\left(\rd_{\ell}\G^{k}_{ij}-\G^{p}_{\ell i}\G^{k}_{pj}\right)du^{\ell}\land du^j\eta_k\cr
&+\left(\rd_m R^{k}_{ij\bar\ell}-\G^{p}_{mj} R^{k}_{ip\bar\ell}-\G^{p}_{mi} R^{k}_{pj\bar\ell}
+\G^{k}_{mp} R^{p}_{ij\bar\ell}-S^{k}_{mji\bar\ell}\right)du^m\psi^{\bar\ell}\rho^j\eta_k\cr
&+\frac{1}{2}\left(\rd_{\bar n}S^{m}_{ijk\bar\ell}-R^{m}_{pi\bar
n}R^{p}_{jk\bar\ell}-R^{m}_{pj\bar n}R^{p}_{ki\bar\ell}
-R^{m}_{pk\bar n}R^{p}_{ij\bar\ell}\right)\psi^{\bar n}\psi^{\bar\ell}\rho^j\land\rho^k\eta_m,\cr
}
$$
The condition $Q^2 H^i=0$ follows from the condition $\bar\rd R=0$, eq.\ \aai,  (note that $\psi^{\bar n}\psi^{\bar\ell}
=-\psi^{\bar \ell}\psi^{\bar n}$).
To show $Q^2\chi_i=0$, we use the condition $[\nabla,\nabla]=0$, eq.\ \aac,
(note that $du^\ell \wedge du^j = -du^j\wedge du^\ell$),
and the definition $S^i{}_{mjk\bar\ell}=\nabla_m R^i{}_{jk\bar\ell}$, eq.\ \aah, as well as the "Bianchi identity",
eq.\ \aaj\ (note that $\r^j\wedge \r^k = \r^k\wedge\r^j$ since $\r^i$ is an anti-commuting $1$-form).

Now it is not difficult to construct our Lagrangian $\sL$ of $U=0$, which is invariant under $Q$ 
and is covariant on $M$. We take
\eqn\aba{
\sL =  Q V,
}
where
\eqn\abb{
V = 
\int_M \biggl( 
\c_i \wedge H^i 
+g_{i\bar j}\r^i\wedge * du^{\bar j}
+ \bar\eta^i *\mathfrak{D}^* w_i
\biggr).
}
We note that $V$ is covariant on $M$ and of $U=-1$. It follows that $\sL=QV$ is
$Q$-invariant and of $U=0$. The explicit form of the Lagrangian $\sL$ is given by
\eqn\abc{
\eqalign{
\sL = \int_M &\left(
g_{i\bar j}du^i\wedge * du^{\bar j}
- \r^i\wedge * \overline D \p^{\bar j}g_{i\bar j}
+\c_i\wedge \left( D \r^i + \Fr{1}{2}R^i{}_{jk\bar\ell}\r^j\wedge\r^k \p^{\bar \ell}\right)^+
\right.
\cr
&
+ H^i\wedge \mathfrak{D}^+ w_i 
+ R^k{}_{ij\bar \ell}\eta_k  H^i\wedge H^j\p^{\bar \ell}
- R^k{}_{ij\bar \ell}\eta_k  H^i\wedge\left(\r^j\wedge du^{\bar\ell}\right)^+
\cr
&
\left.
-\Fr{1}{2}S^{n}_{ijk\bar \ell}\eta_n  H^i\wedge \left(\r^j\wedge\r^k\right)^+ \p^{\bar \ell}
+v^i  *\mathfrak{D}^* w_i + \bar\eta^i  *\mathfrak{D}^*\mathfrak{D}\eta_i
\right).
}
}


\subsection{An Analogy with Topological String B Model}

Now we want to discuss an analogy with the Topological String B model \cite{W1}.
Let $\S$ be a smooth oriented compact Riemannian surface. 
We denote the space of differential $r$-form on $\S$ by $\O^r(\S)$ and the
the exterior derivative on $\S$  also by $d$. 
We pick a Riemannian metric $h$ on $\S$ and denote $*$ the Hodge star operator on $\S$.
We first note that the space $2$-form on $\S$ is isomorphic the space of $0$-form.

The B model  governs maps
$\phi: \S \rightarrow X$ with the choice of a Riemannian metric $h$ on $\S$,
a K\"aher metric $g_{i\bar j}$ on $X$ and a complex structure $J$ on $X$.
We adopt the same conventions involving the target Calabi-Yau space $X$ as in our model.
Picking a local holomorphic coordinates $\{u^i\}$,
a map $\phi$ can be described locally via  local coordinates $(u^i, u^{\bar i})$  on $X$ regarded as functions on $\S$. 
We have an integral valued quantum number called ghost number
$U$ and assign  $U=0$ to the bosonic fields $(u^i, u^{\bar i})$.
We have  an anti-commuting $0$-form field $\p^{\bar i}$ of $U=1$, a section of $\phi^*(\overline\CT)$,
the pullback of anti-holomorphic tangent bundle of $X$ to $M$. 
We also have an anti-commuting $1$-form field $\r^i$ of $U=-1$,  a section of $\phi^*(\CT)$. 
Finally we have a anti-commuting $0$-form field $\chi_i$ of $U=1$, a section of $\phi^*(\CT^*)$.
The fermionic  transformation laws are
\eqn\gabd{
\eqalign{
&Q u^i =0
,\cr
&Q \r^i =du^i,
}
\qquad
\eqalign{
Q u^{\bar i} &=\p^{\bar i},\cr
Q\p^{\bar i} &=0,\cr
}
\qquad Q \chi_i=0.
}
The global fermionic supercharge $Q$ carries $U=1$ and
satisfies $Q^2=0$. 
The Lagrangian  is given by
\eqn\gabc{
\eqalign{
\sL_B = \int_\S &\left(
g_{i\bar j}du^i\wedge * du^{\bar j}
- \r^i\wedge * \overline D \p^{\bar j}g_{i\bar j}
+\c_i\left( D \r^i + \Fr{1}{2}R^i{}_{jk\bar\ell}\r^j\wedge\r^k \p^{\bar \ell}\right)
\right).
}
}

We note that the $0$-form field $\chi_i$ on $\S$ can be dualized to a $2$-form field on $\S$.
Now a straightforward generalization to $4$-dimensional sigma model might be obtained by
declaring $\chi_i$ to be a self-dual $2$-form field on  $4$-manifold $M$ while maintaining the form degrees and
ghost numbers of the all the other fields;
$$
\eqalign{
\sL^\pr = \int_M &\left(
g_{i\bar j}du^i\wedge * du^{\bar j}
- \r^i\wedge * \overline D \p^{\bar j}g_{i\bar j}
+\c_i\wedge\left( D \r^i + \Fr{1}{2}R^i{}_{jk\bar\ell}\r^j\wedge\r^k \p^{\bar \ell}\right)^+
\right).
}
$$
An obvious problem of the above Lagrangian is that the field $\chi_i$, now a self-dual $2$-form on $M$ has 
a new gauge degrees of freedom.  
Our Lagrangian $\sL$ in \abc\ may viewed as a result of removing such degeneracy as follows.
The additional $1$-form field
$w_i$ with $U=2$ is interpreted as the ghost for the  new gauge degree of freedom of the self-dual $2$-form field $\chi_i$ of $U=1$,  while the self-dual $2$-form field $H^i$ with $U=-2$ is an auxiliary field for its gauge fixing - as is shown in
the term $H^i\wedge \mD^+ w_i$ in our Lagrangian $\sL$.
Being an $1$-form field  $w_i$ has residual gauge degree of freedom
and the additional $0$-form field $\eta_i$ with $U=3$ is   the ghost for the residual gauge degree of freedom,
while the additional $0$-form field $\bar\eta^i$ with $U=-3$ is the antighost of the ghost $\eta^i$
and the additional $1$-form field $v^i$ with $U=-2$ is
the auxiliary field for its gauge fixing - as is shown in
the terms $v^i*\mD^* w_i + \bar \eta^i *\mD^* \mD\eta_i$ in our Lagrangian $\sL$.

Besides from the above formal comparison our model will be shown to have those key properties of the topological string
B model that the theory is independent of both the Riemannian metric $h$ on $\S$ and the  K\"ahler metric $g_{i\bar j}$ on $X$
but depends on the complex structure on the target space $X$.

\subsection{Two Important Properties}

Our Lagrangian $\sL$ in \abc\ depends on the Riemannian metric $h_{\a\b}$ 
on $M$ via the associated Hodge star $*$ operator,
the K\"ahler metric $g_{i\bar j}$ on $X$, as well as the complex structure $J$ on $X$.   
In this subsection we will show that the infinitesimal variations of $\sL$ with respect to both 
the Riemannian metric $h_{\a\b}$ on $M$ and the K\"ahler metric $g_{i\bar j}$ on $X$ are $Q$-exact
so that the partition function of the theory and the expectation values of $Q$-invariant observables  depend 
only on the complex structure on $X$. In particular the partition function and  the expectation values of 
$Q$-invariant observables are, following Witten's arguments in Donaldson-Witten theory \cite{W2},  smooth invariants of $4$-manifold.

We first establish that the ($4$-dimensional) energy-momentum tensor $T_{\a\b}$ is $Q$-exact.
$T_{\a\b}$ is defined in terms of the variation of the Lagrangian with respect to an infinitesimal change of
the $4$-dimensional Riemannian metric $h^{\a\b}\longrightarrow h^{\a\b} + \d h^{\a\b}$ such that
\eqn\emta{
\d \sL = \Fr{1}{2}\int_M \sqrt{h} \d h^{\a\b}T_{\a\b},
}
where $h$ denotes the determinant of $h_{\a\b}$. The computation is straightforward besides from a slight subtlety 
that the two  $2$-form fields $\chi_i=\fr{1}{2}\chi_{i\a\b}dx^\a\wedge dx^\b$
and $H^i=\fr{1}{2}H^i_{\a\b}dx^\a\wedge dx^\b$ among the all fields are subjected to the self-duality conditions
\zabn.
Thus the self-duality of those fields should be preserved when computing the variation \emta, as  in the similar situation
of  Donaldson-Witten theory. So an infinitesimal change $\d h^{\a\b}$ must be accompanied by the variations
\eqn\emtb{
\eqalign{
\d H^i_{\a\b} &=\frac{1}{2}\e_{\a\b\g\d}\d h^{\g\g'}h^{\d\d'} H_{\g'\d'}^{i}
-\frac{1}{8}(\d h^{\s\t}h_{\s\tau})\e_{\a\b\g\d}\d h^{\g\g'}h^{\d\d'} H^{i}_{\g'\d'}
,\cr
\d \chi_{i\a\b} &=\frac{1}{2}\e_{\a\b\g\d}\d h^{\g\g'}h^{\d\d'} \chi_{i\g'\d'}
-\frac{1}{8}(\d h^{\s\t} h_{\s\tau})\e_{\a\b\g\d}\d h^{\g\g'}h^{\d\d'} \chi_{i\g'\d'}.
}
}
Then a tedious but straightforward computation gives
\eqn\energy{
\eqalign{
T_{\a\b}=&
2\left(\rd_{\a}u^i\rd_{\b}u^{\bar{j}}g_{ij}
-\rho_{\a}^i\bar{D}_{\b}\psi^{\bar{j}}g_{i\bar{j}}
+\lambda^i\mD_{\a} w_{i\b}
+\bar{\eta}^i\mD_{\a}\mD_{\b}\eta_i\right)
\cr
&-h_{\a\b}\left(
g_{i\bar{j}}\rd_{\g}u^i\rd^{\g}u^{\bar{j}}
-\rho_{\g}^{i}\bar{D}^{\g}\psi^{\bar{j}}g_{i\bar{j}}
+\lambda^i\mD_{\g} A_{i}^{\g}
+\bar{\eta}^i\mD_{\g}\mD^{\g}\eta_i
\right)
\cr
&+2 \chi_{i\a\g}\left(
D_{\b}\rho^{i\g}
-D^{\g}\rho_{\b}^{i}
+R_{jk\bar\ell}^{i}\rho_{\a}^{j} \rho_{\b}^{k}\psi^{\bar{\ell}}
\right)
\cr
&-h_{\a\b} \chi_{i\g\d}\left(
D^{\g}\rho^{i\d}
-D^{\d}\rho^{i\g}
+R_{jk\bar{\ell}}^{i}\rho^{j\g} \rho^{k\d}\psi^{\bar{\ell}}
\right)
\cr
&+2\biggl(\mD_{\a} \w_{i\g}
-\mD_{\g} \w_{i\a}
+R_{ij\bar{\ell}}^{k}\eta_{k} H_{\a\g}^j\psi^{\bar{\ell}}
-S^{n}{}_{ijk\bar{\ell}}\eta_{n}\rho_{\a}^{j}\rho_{\g}^{k}\psi^{\bar{\ell}}
\cr
&
\qquad+R^{k}{}_{ij\bar{\ell}}\eta_{k}\left(\rd_{\a}u^{\bar{\ell}}\rho_{\g}^j-\rd_{\g} u^{\bar{\ell}}\rho_{\a}^j\right)
\biggr)H_{\b}^{i\g}
\cr
&-h_{\a\b}\biggl(
\mD_{\g} w_{i\d}-\mD_{\d} w_{i\g}
+R^k{}_{ij\bar{\ell}} H_{\g\d}^{j}\psi^{\bar{\ell}}
-S^n{}_{ijk\bar{\ell}}\eta_{n}\rho_{\g}^{j}\rho_{\d}^{k}\psi^{\bar{\ell}}
\cr
&\qquad +R^k{}_{ij\bar{\ell}}\eta_{k}\left(\rd_{\g} u^{\bar{\ell}}\rho_{\d}^j-\rd_{\d} u^{\bar{\ell}}\rho_{\g}^j\right)
\biggr)H^{i\d\g}
}
}

We note that the variations \emtb\ commutes with $Q$ since  $Q$ transforms as a scalar  in $4$-dimensions,
so that it does not depend on the Riemannian metric and the application of $Q$ to a self-dual
$2$-form gives a self-dual $2$-form on $M$. That is
$$
\eqalign{
Q\d H^i_{\a\b} &=\frac{1}{2}\e_{\a\b\g\d}\d h^{\g\g'}h^{\d\d'} Q H_{\g'\d'}^{i}
-\frac{1}{8}(\d h^{\s\t}h_{\s\tau})\e_{\a\b\g\d}\d h^{\g\g'}h^{\d\d'} Q H^{i}_{\g'\d'}
\cr
 &=\d QH^i_{\a\b},
}
$$
and $Q\d \chi_{i\a\b}=\d Q\chi_{i\a\b}$ by the same reason.
All the other fields, being either $0$-form or $1$-form fields, do not depend on metric on $M$.
Thus we have
$[Q,\d](any fields)=0$.
Recalling that our Lagrangian is  given by $\sL=Q V$, 
we define $\l_{\a\b}$ in terms of the variation of $V$ \aab\  with respect to  
$h^{\a\b}\longrightarrow h^{\a\b} + \d h^{\a\b}$ such that
\eqn\emtc{
\d V = \Fr{1}{2}\int_M \sqrt{h} \d h^{\a\b}\l_{\a\b}.
}
{}From $\d\sL \equiv \d QV=Q\d V$ we conclude that
\eqn\cruca{
T_{\a\b} = Q \l_{\a\b}.
}
A direct computation $\d V$ with respect to $h^{\a\b}\rightarrow h^{\a\b} + \d h^{\a\b}$  
accompanying the variations \emtb\
gives us 
\eqn\zaca{
\eqalign{
\l_{\a\b}=
& \chi_{i\a\g}  H_{\b}^{i\g}
+ \chi_{i\b\g} H_{\a}^{i\g}
-\fr{1}{2}h_{\a\b} \chi_{i\g\d} H^{i\g\d},
\cr
&+\bar{\eta}^i\mD_{\a} w_{i\b}+\bar{\eta}^i\mD_{\b}
w_{i\a}
-\bar{\eta}^i\mD_{\g} w_{i}^{\g}
\cr
&+\rho_{\a}^i \cdot \rd_{\b}u^{\bar{j}}
g_{i\bar{j}}+\rho_{\b}^i \cdot \rd_{\a}u^{\bar{j}}g_{i\bar{j}}
-h_{\a\b}\rho_{\g}^i \cdot \rd^{\g}u^{\bar{j}} g_{i\bar{j}}.
}
}
One can also check explicitly that $T_{\a\b} = Q\l_{\a\b}$.

Now we want to establish that our theory is independent of the K\"{a}hler metric of the target space $X$.
For this, we study the variation $\d \sL$ of the Lagrangian $\sL$ with respect to an infinitesimal change
of the K\"ahler metric $g^{i\bar j}\rightarrow g^{i\bar j} + \d g^{i\bar j}$ preserving the K\"ahlerian condition.
Such a variation is accompanied with the following induced infinitesimal variations
\eqn\abk{
\eqalign{
\d \G^i{}_{j k}&=g^{\bar n i}\rd_j \d g_{\bar n k} + \d g^{\bar n i}\rd_j g_{\bar n k}
,\cr
\d R^i{}_{jk\bar \ell}&=\rd_{\bar \ell}\d\G^i{}_{jk}
,\cr
\d S^i{}_{jk\ell\bar m}&=\nabla_j\left(\rd_{\bar m} \d\G^i{}_{k\ell}\right) 
- \d\G^{i}{}_{jn}R^{n}{}_{k\ell \bar m}
- \d\G^n{}_{jk}R^i{}_{n\ell\bar m}.
}
}
It may not be obvious if such variation $\d \sL$ is $Q$-exact since
the fermionic transformation laws for the fields $H^i$, $w_i$ and $\chi_i$
depend on the K\"{a}hler metric $g_{i\bar j}$ on $X$. But a direct computation shows that
\eqn\abki{
\eqalign{
\d \sL =Q\int_M\biggl(&-\frac{1}{2}\d\G^{i}_{jk}\chi_i\wedge\rho^j\wedge\rho^k+\d\G^{k}_{ij}H^i\wedge A_k\land\rho^j
-\d\G^{k}_{ij}\eta_k H^i\land H^j 
\cr
&+\frac{1}{2}H^i\wedge\nabla_i(\d\G^{n}_{jk})\eta_n\rho^j\wedge\rho^k+\d g_{i\bar j}\rho^i\wedge *du^{\bar j}
\cr
&-\d\G^{k}_{ij}\bar{\eta}^i du^j\wedge *A_k
-\rd_{\bar\ell}(\d\G^{k}_{ij})\bar{\eta}^i \psi^{\bar\ell}\rho^j\wedge *A_k \biggr),
}
}
under $g^{i\bar j}\rightarrow g^{i\bar j} + \d g^{i\bar j}$.

Similarly to the B model in two dimensions, our theory
obviously depends on the variation
of complex structures on the target space $X$.
Then,  moduli space $\CM(X)$ of complex structures on $X$ becomes the
moduli space of the theory. Thus we may obtain family of
smooth invariants of the four manifold  parametrized by the moduli space
of complex structures of $X$.


%

\subsection{Localization and Fermionic Zero Modes}


Rescaling $\sL$ by $t\sL$,
the relation \cruca\ also implies that the theory is independent of $t$ as long as
$\Re t > 0$.
Thus we may take the semi-classical limit  $\Re t\rightarrow \infty$
such that the path integral has dominant contribution from the
minima of the bosonic part of the action functional. Such minima
are the constant maps $du^i =0$. Thus the path integral is
localized to the space of constant maps, which is a copy of
the target manifold $X$ itself. Then one may perform perturbative
calculation after expanding around constant maps, while
taking care of bosonic and fermionic zero-modes.
In the one loop computation we will encounter with
fermion determinant. We note that the zero forms $\p^{\bar i}$ and
self-dual two-forms $g^{\bar j i}\c_i$ are sections of $\bar\CT=T^{0,1}_X$,
while the one-forms $\r^i$ are sections of $\CT =T^{1,0}_X$.
Consequently the fermion determinant is complex.
The situation is similar to the B model in two dimensions
and the condition to have well-defined fermion determinant
is that $c_1(X)=0$, $X$ is a Calabi-Yau manifold.

Expanding around the constant map $M\rightarrow X$ 
the zero modes of the fermions $\p^{\bar i}$ with $U=1$ are the constant modes
of $\p^{\bar i}$, i.e., $d\p^{\bar i}$, the zero-modes of $\r^i$ with $U=-1$ are
the harmonic one-forms, i.e., $d^+\r^i = d*\r^i =0$ and the
zero modes of $\chi_i$ with $U=1$ are the self-dual harmonic two-forms, i.e., 
$d^{*+} \c_i =0$. Thus those zero-modes are harmonic representatives of cohomology classes
of the  following complex
\eqn\abh{
0 \longrightarrow \O^0(M)\ato{d} \O^1(M)\ato{d^+}\O^2_{+}(M)\rightarrow 0,
}
where $d^+ = P^+\circ d$ and $P^+:\O^2(M)\rightarrow \O^2_+(M)$.
Note that the zero-modes of $\eta_i$ with $U=3$ and the zero-modes of $\bar\eta^i$ with $U=-3$
always appear in the pairs.
Thus the ghost number anomaly (the net violation of the ghost
number in the path integral measure due to the fermionic zero-modes)
are
\eqn\abi{
\# U = d (1 - b_1(M) + b^2_+(M))= \Fr{d}{2}\left(\c(M) +\s(M)\right)
}  
where $b_1(M)$ denotes the first Betti number of $M$ and $b^2_+(M)$
the number of the positive eigen values in the intersection matrix
of $H_2(M)$, $\c(M) = 2(1 -b_1(M)) + b_2(M)$ is the Euler number of $M$ and 
$\s(M)= b^2_+(M) -b^2_-(M)$ is
the signature. 
It follows that the partition function of the theory is well-defined
if and only if the four manifold $M$ has
$\c(M) + \s(M) =0$.
Otherwise we have to insert suitable combination of observables
to have non-vanishing correlation functions.

\subsection{The Observables}

Now we discuss the observables of the theory, which are defined as representatives of $Q$-cohomology classes. 
The situation will be quite similar to that of topological string B model in the paper \cite{W1}, and 
we will follow Witten's discussions there closely. 

Consider the
following transformation laws from \abd\ and \abda;
\eqn\aca{
\eqalign{
Q u^i &=0,\cr
Q \eta_i &=0,
}\qquad\eqalign{
Q u^{\bar i} &= \p^{\bar i},\cr
Q \p^{\bar i} &=0,\cr
}
}
where the field $\p^{\bar i}$ is a section of $\phi^*(\overline\CT)$ of $U=1$
and the field $\eta_i$ is a section of $\phi^*(\CT^*)$ of $U=3$. 
On the target space $X$ we have the Dolbeault complex $( \O^{0,\bullet} (X, \wedge^\bullet \CT),\bar\rd)$
for the sheaf cohomology group $H^{0,\bullet}_{\bar \rd} (X, \wedge^\bullet \CT)$.
For any $(0,p)$-form $W$ on $X$ with value in $\wedge^q \CT$,
$W \in \O^{0,p} (X, \wedge^q \CT)$ we associate an operator 
\eqn\acb{
\CO^{(0)}_W = \Fr{1}{p! q!}\p^{\bar j_1}\ldots \p^{\bar j_p}W_{\bar j_1\ldots \bar j_p}{}^{i_1\ldots i_q}\eta_{i_1}\ldots\eta_{i_q}
}
with the ghost number $U=p + 3 q$
Then we find that
\eqn\acc{
Q\CO^{(0)}_W = \CO^{(0)}_{\bar\rd W}.
}
Thus for any representative $W$ of  the sheaf cohomology class 
$[W]\in H^{0,p}_{\bar \rd} (X, \wedge^q \CT)$ we have an observable $\CO^{(0)}_W$
of $U=p+3q$, which is a representative of non-trivial $Q$-cohomology class.
Such observables are $0$-forms on $M$ and to be inserted on points in $M$.
The analogy with the topological string $B$-model is obvious.

 \newsec{Toward the Topological Family}
 
 In this section we discuss the possibility of constructing more general family of topological Lagrangians.
 
The basic idea is very simple; assume that there is a representative  of $Q$-cohomology class, which is
given by integral of a $4$-form $\CO^{(4)}$ on $M$ over $M$, i.e., $\int_M \CO^{(4)}$, then we may get
an one parameter family of topological Lagrangians,
$$
\sL \longrightarrow \sL + t \int_M \CO^{(4)}
$$
invariant under $Q$. Obviously we may also consider multiple parameters family by including every $Q$-cohomology
class, which representative is the similar integral over $M$. 
The standard procedure for finding such cohomology classes is using so called the descent equations. 
To begin with we fix a homogeneous basis $\{[\g_a]\}$ of $H^{0,\bullet}_{\bar \rd} (X, \wedge^\bullet \CT)$.
Let $\{\g_a\}$ be a set of representatives and let $\{\CO^{(0)}_a\}$ be the associated set of representatives
of $Q$-cohomology classes which are $0$-forms on $M$.
Then one wants  to solve the following sequence of equations  called the descent equations;
\eqn\acd{
\eqalign{
0&=Q \CO^{(0)}_a,\cr
d\CO^{(0)}_a &= Q\CO^{(1)}_a,\cr
d\CO^{(1)}_a &= Q \CO^{(2)}_a,\cr
d\CO^{(2)}_a &= Q \CO^{(3)}_a,\cr
d\CO^{(3)}_a &= Q \CO^{(4)}_a,\cr
}}
where $\CO^{(r)}_a$ is an $r$-form on $M$ with the ghost number 
$U (\CO^{(r)}_a)=U (\CO^{(0)}_a)  -r$. 
Let's assume, {\it temporarily}, that we have solutions of the above set of equations.
Then, the second equation implies that the path integral of observable $\CO^{(0)}_a$
does not depend on its position at $M$, to where $\CO^{(0)}_a$ is inserted.
In general we pick an $r$-cycle $C_{(r)}$ on $M$ and consider $\int_{C_{(r)}}\CO^{(r)}_a$.
{}From \acd, we have
\eqn\ace{
Q \int_{C_{(r)}}\CO^{(r)}_a= \int_{C_{(r)}} d \CO^{(r-1)}_a=  \int_{\rd C_{(r)}} \CO^{(r-1)}_a,
}
which relation means that  the $Q$ cohomology of $\int_{C_{(r)}}\CO^{(r)}_a$
depends only on the homology of $C_{(r)}$. In particular for a homology cycle $C_{(r)}$,
$\int_{C_{(r)}}\CO^{(r)}_a$ gives  a non-trivial $Q$-cohomology class.
Furthermore one may add $\int_M \CO^{(4)}_a$ to the original Lagrangian
$\sL$ and get a family of topological Lagrangian.  For this we may introduce
graded parameters $\{t^\a\}$ regarded as the dual basis of
$H^{0,\bullet}_{\bar \rd} (X, \wedge^\bullet \CT)$ - dual to the homogeneous basis $\{[\g_\a]\}$,
after shifting ghost number by $4$ such that $U(t^a) + U(\CO^{(4)}_a)=0$. 
Then the desired family of Lagrangian $\sL_t$ is
\eqn\acea{
\sL_t = \sL +  t^a \int_M \CO^{(4)}_a,
} 
of $U=0$.  
We shall, however, see that we can not solve the descent equations completely.

It is rather straightforward to find $\CO^{(1)}_a$ solving the second equation in \acd.
We first note that 
\eqn\acg{
d \CO^{(0)}_a = du^i \Fr{\d \CO^{(0)}_a}{\d u^i} 
+ d\eta_i \Fr{\d \CO^{(0)}_a}{\d \eta_i}
+ d\p^{\bar i} \Fr{\d \CO^{(0)}_a}{\d \p^{\bar i}}.
}
We also note that
the fermionic transformation laws \abd\ and \abda\ imply  the following relations
\eqn\acf{
Q\r^i=du^i,\quad Q(w_i + \G^k{}_{ij}\eta_k \r^j)= - d\eta_i,\quad Q du^{\bar i}= d\p^{\bar i},
}
as well as
\eqn\acfa{
\left[Q, \Fr{\d}{\d u^i}\right]=0,\quad
\left[Q, \Fr{\d}{\d u^{\bar i}}\right]=0,\quad 
\left\{Q, \Fr{\d}{\d \eta^i}\right\}=0,\quad 
\left\{Q, \Fr{\d}{\d \p^{\bar i}}\right\}=\Fr{\d}{\d u^{\bar i}}.
}
Now using $Q \CO^{(0)}_a =0$ and  the relations \acf\ and \acfa\
we deduce that $\CO^{(1)}_a$ given by
\eqn\ach{
\CO^{(1)}_a = 
\r^i\Fr{\d\CO^{(0)}_a}{\d u^i} 
-(w_i + \G^k{}_{ij}\eta_k \r^j)\Fr{\d\CO^{(0)}_a}{\d \eta_i}
+ du^{\bar i}\Fr{\d\CO^{(0)}_a}{\d \p^{\bar i}},
}
satisfies $ Q\CO^{(1)}_\a = d \CO^{(0)}_\a$.
One may proceed to find $\CO^{(2)}_a$ in the similar fashion. After a tedious but straightforward
computations we obtain 
the following expression
\eqn\acf{
\eqalign{
\CO^{(2)}_a = & 
-\left(H^i +\fr{1}{2}\G^i{}_{jk} (\r^j\wedge \r^k)^+\right) \Fr{\d\CO^{(0)}_a}{\d u^i}
\cr
&
-\left(\chi_i -\G^k{}_{ij}(w_k\wedge\r^j)^+ +\G^k{}_{ij}\eta_k H^j - \fr{1}{2}\rd_i \G^\ell{}_{jk}\eta_\ell(\r^j\wedge\r^k)^+\right)
\Fr{\d\CO^{(0)}_a}{\d \eta_i}
\cr
&
+\Fr{1}{2}\r^i\wedge\r^j \Fr{\d^2\CO^{(0)}_a}{\d u^j \d u^i}
+\Fr{1}{2}\left(w_i+\G^k{}_{i m}\eta_k \r^m\right)\wedge \left( w_j+\G^\ell{}_{j n}\eta_\ell \r^n \right)\Fr{\d^2\CO^{(0)}_a}{\d \eta_j \d \eta_i}
\cr
&
+\Fr{1}{2}du^{\bar i}\wedge du^{\bar j}\Fr{\d^2\CO^{(0)}_a}{\d \p^{\bar j} \d \p^{\bar i}}
+\r^i\wedge \left( w_j+\G^\ell{}_{j n}\eta_\ell \r^n \right)  \Fr{\d^2\CO^{(0)}_a}{\d \eta_j \d u^i}
\cr
&
+\r^i\wedge du^{\bar j}  \Fr{\d^2\CO^{(0)}_a}{\d \p^{\bar j}\d u^i}
+\r^i\wedge du^{\bar j}  \Fr{\d^2\CO^{(0)}_a}{\d \p^{\bar j}\d\eta_i},
}
}
satisfies the relation $ Q\CO^{(2)}_a = d \CO^{(1)}_a$. 

It turns out that a solution
of the next descent equation $d\CO^{(2)}_a = Q\CO^{(3)}_a$ for  $\CO^{(3)}_a$
does not exist in the strict sense. There is, however, a solution $\CO^{(3)}_a$ of the descent equation
modulo the equations of motion;
\eqn\acg{
d \CO^{(2)}_a = Q \CO^{(3)}_a + G_a,
}
where $G_a$ is certain expression that vanishes modulo the $H_i$ and $\c_i$ equations of motion.\foot{The precise 
expressions for $\CO^{(3)}_a$ and $G_a$, which are very complicated and unilluminating in the face values, are beyond the purpose of this paper. It is, however,
It is not difficult to see what is the origin of the above.
Looking back the transformation laws \abd, we notice that
\eqn\ach{
\eqalign{
Q u^i &=0,\cr
Q\eta_i &=0,\cr
}\qquad
\eqalign{
Q \r^i &=du^i,\cr
Q w_i &=-d\eta_i +\cdots,\cr
}\qquad
\eqalign{
Q H^i &=-d^+\r^i+\cdots,\cr
Q \c_i &=d^+w_i +\cdots,\cr
}
}
while we do not have $3$-form fields, which the fermionic transformation
laws contain $d H^i$ and $d \c_i$ terms.
}

The similar phenomena had been first observed  in the topological string  B model \cite{W1}.
Adopting Witten's recipe there
we may  proceed further to find $\CO^{(4)}_a$ satisfying the descent equation
modulo equations of motion;
\eqn\acg{
d\CO^{(3)}_a = Q\CO^{(4)}_a
 + \sum_A \Fr{\d \sL}{\d \phi_A}\cdot \zeta_{a,A},
}
where $\phi_A$ are all the fields of the theory. In the above 
$\Fr{\d \sL}{\d \phi_A}$ are equations of motion, so the $2$-nd terms in the right hand side of \acg\
vanishes   by the equation. 
The above relation implies that the first oder deformation toward the topological family 
\eqn\acea{
\sL^\pr = \sL + t^a \int_M \CO^{(4)}_a.
} 
may not be invariant under the original $Q$, but is invariant under the following 
new $Q^\pr$ 
\eqn\acab{
Q^\pr \phi_A = Q \phi_A +  t^a \zeta_{a,A}
} 
up to the terms in the second order in $t^a$. The new BRST charge $Q^\pr$ is also nilpotent only up to the second order.
Thus we have to repeat the whole procedure again and again
until we fill up all the higher order corrections to both the  topological Lagrangian
and the BRST charge
\eqn\aceaf{
\eqalign{
\sL_t&= \sL + t^a \int_M \CO^{(4)}_a +  \sum_{\ell=2}^\infty  t^{a_1}\cdots t^{a_\ell} \int_M \CO^{(4)}_{a_1\cdots a_\ell},
\cr
Q_t \phi_A &= Q \phi_A +  t^a \zeta_{a,A}+\sum_{\ell=2}^\infty  t^{a_1}\cdots t^{a_\ell}\zeta_{a_1\cdots a_\ell,A},
} 
}
such that $Q_t \sL_t=0$ and $Q_t^2=0$.
Assuming that such all order corrections are possible,
the above discussion implies that we have certain extended moduli space 
$\CN$ of the theory such that tangent space to $\CN$ at the origin
is 
\eqn\acae{
T\CN|_o \simeq \bigoplus_{p,q=0}^d H^{0,p}_{\bar\rd}(X, \wedge^q \CT),
}
-the space of all $Q$ cohomology of zero-dimensional
observables. Then the remaining problem
is to determine $\CN$ and construct the family of topological theories parametrized by $\CN$.

We should remark again that the situation is closely related to that of  the B model in
two-dimensions \cite{W1}. We note that the tangent space \acae\ 
to the moduli space $\CN$ is exactly same with the tangent space
of the thickened moduli space at a classical point as discussed by Witten for
the B model in two dimensions. We claim that  the two moduli spaces
are  isomorphic to  the extended moduli space
of complex structures of the Calabi-Yau $X$, defined by Barannikov and Kontsevich \cite{BK}.
The second named author showed that the extended moduli space of complex structure is the thickennd moduli space 
parametrizing the family of B-models  by actually constructing such a  family in Section $4.2$ of \cite{P1}.

The solution of our problem should be obtained by going beyond
the on-shell and adopting more powerful viewpoints as was advocated in the papers \cite{P1,P2}.
The basic strategy is to recast our original Lagrangian $\sL$ as a gauge fixed version of certain
master action functional $S$, adopting the Batalin-Vilkovisky quantization scheme \cite{BV,AKSZ}. 
Then our model
is interpreted as a theory of maps  
$$\Phi:T[1]M\longrightarrow\overline\CT[1]\oplus \CT^*[3],$$
where $T[1]M$ is the superspace obtained from the total space of tangent bundle to $M$ after 
twisting the fiber by $U=1$, $\overline\CT[1]$ be the total space of the anti-holomorphic tangent bundle $\overline\CT= T^{0,1}X$
to the Calabi-Yau space $X$ after twisting the fiber by $U=1$ and $\CT^*[3]$ be the total space of the holomorphic cotangent bundle $\CT^*= T^{*1,0}X$ to $X$ after twisting the fiber by $U=3$. 
Then one can construct a BV action functional $S$ from a smooth structure on $M$ and the complex structure on $X$,
satisfying  so called the classical master  equation $(S,S)=0$, such that our Lagrangian $\sL$ is obtained after gauge fixing $\sL = S^{g.f.}$ and the BRST charge $Q$ is give by $Q\phi_A = (S, \phi_A)^{g.f.}$

The standard description of the extended moduli space  $\CN$ of complex structures \cite{BK} shall be rephrased by
extended moduli space associated with the natural differential graded Lie $3$-algebra $(\mt,\bar\rd, [\bullet,\bullet])$, 
where  $\mt=\bigoplus_{r=0}^{4n}$ the space of functions on $\overline\CT[1]\oplus \CT^*[3]$ - isomorphic to 
$$
\mt^r\simeq \bigoplus_{r=p+3q} \O^{0,p}(X, \wedge^q \CT),
$$
$\bar\rd$ is the  Dolbeault operator for the given complex structure - regarded as an odd vector field of $U=1$
on $\overline\CT[1]\oplus \CT^*[3]$, and the bracket $[\bullet,\bullet]$ is the graded Poisson bracket of $U=-3$ 
associated with $\CT^*[3]$ extended naturally to $\mt$. Let $\mh$ be the cohomology of the complex $(\mt,\bar\rd)$ and let $[\g_\a]$ be a homogenous basis of $\mh$. Let $\{t^\a\}$ be dual to $[\g_\a]$ after shifting $U$ by $4$ in the sense that
$U(t^\a)= - U([\g_\a]) + 4$. Now consider another differential graded Lie $3$-algebra $(\Bbbk[[t^\a]]\otimes \mt, \bos{1}\otimes \rd, [\bullet,\bullet])$.
Then a result of \cite{BK} also implies that we have a versal solution
$\G \in (\Bbbk[[t^\a]]\otimes \mt)^4$ to the Maurer-Cartan equation 
\eqn\fya{
\bar\rd \G +\Fr{1}{2}[\G,\G]=0,
}
in the
form
\eqn\fyb{
\G = t^a \g_a +\sum_{\ell=2}^\infty \Fr{1}{\ell !}t^{a_1}\cdots t^{a_\ell} \g_{a_1\cdots a_\ell}
}
where $\{\g_a\}$ is a set of representative of $\{[\g_a]\}$ and $\g_{a_1\cdots a_n} \in \mt^{\sum_{k=1}^\ell U(\g_{a_k}) - 4\ell}$.
The Maurer-Cartan equation \fya\ implies that $\bar\rd_\G := \bar\rd + [\G,\hbox{ }]$ satisfies $\bar\rd_\G^2=0$.
We claim that the associated BV action functional $S_\G$ defined as follows
\eqn\fyc{
\eqalign{
S_\G:=& S + \int_{T[1]M}\Phi^*(\G)
\cr
=& S + t^\a\int_{T[1]M}\Phi^*(\g_\a)+\sum_{\ell=2}^\infty \Fr{1}{\ell !}t^{a_1}\cdots t^{a_\ell}\int_{T[1]M}\Phi^*(\g_{a_1\cdots a_\ell}) ,
}
}
where $\int_{T[1]M}$ is the superspace integration over $T[1]M$,
also satisfies the classical BV master equation  
\eqn\cbhu{
(S_\G,S_\G)=0.
}

Our claim is, then, $S_\G$  gives both the desired  topological Lagrangian
$\sL_t$ and the BRST charge $Q_t$ in the form of \aceaf\ parametrized by $\CN$ after a gauge fixing
- i.e.,  $\sL_t=(S_\G)^{g.f.}$ such that $S^{g.f.}=\sL$ and
$\left(\int_{T[1]M}\Phi^*(\g_{a_1\cdots a_\ell})\right)^{g.f.}= \int_M \CO^{(4)}_{a_1\cdots a_\ell}$
for all $\ell=1,2,3,\ldots$ and  $Q_t \phi_A = (S_\G, \phi_A)^{g.f.}$. Then
the desired relations that $Q_t^2=0$ and  $Q_t \sL_t=0$ are satisfied 
as the results of the BV master equation \cbhu.
Furthermore this approach also leads to stream lined Hamiltonian formalism of the
family of theory as was sketched in Sect. $4.1$ of \cite{P2} for general case. 
The details of these discussions shall appear
elsewhere \cite{JP2}.

\acknowledgments

We would like to thank Hoil Kim for useful discussions. We also like to thank the referee whose comments
helped us  improve the presentation of this paper.

\end{document}